 \documentclass[final,3p,times,sort&compress]{elsarticle}

\usepackage{multirow,setspace,times,amssymb,amsmath,graphicx,color,rotating,subfigure,url}
\usepackage{lineno}
\usepackage{natbib}

\journal{Physica A}

\begin{document}

\begin{frontmatter}


\title{Long-term correlations and multifractal nature in the intertrade durations of a liquid Chinese stock and its warrant}
\author[BS,RCE]{Yong-Ping Ruan}
\author[BS,RCE,SS]{Wei-Xing Zhou \corref{cor}}
\cortext[cor]{Corresponding author. Address: 130 Meilong Road, P.O. Box 114, School of Business,
              East China University of Science and Technology, Shanghai 200237, China, Phone: +86 21 64250053, Fax: +86 21 64253152.}
\ead{wxzhou@ecust.edu.cn} %

\address[BS]{School of Business, East China University of Science and Technology, Shanghai 200237, China}
\address[RCE]{Research Center for Econophysics, East China University of Science and Technology, Shanghai 200237, China}
\address[SS]{School of Science, East China University of Science and Technology, Shanghai 200237, China}

\begin{abstract}
Intertrade duration of equities is an important financial measure characterizing the trading activities, which is defined as the waiting time between successive trades of an equity. Using the ultrahigh-frequency data of a liquid Chinese stock and its associated warrant, we perform a comparative investigation of the statistical properties of their intertrade duration time series. The distributions of the two equities can be better described by the shifted power-law form than the Weibull and their scaled distributions do not collapse onto a single curve. Although the intertrade durations of the two equities have very different magnitude, their intraday patterns exhibit very similar shapes. Both detrended fluctuation analysis (DFA) and detrending moving average analysis (DMA) show that the 1-min intertrade duration time series of the two equities are strongly correlated. In addition, both multifractal detrended fluctuation analysis (MFDFA) and multifractal detrending moving average analysis (MFDMA) unveil that the 1-min intertrade durations possess multifractal nature. However, the difference between the two singularity spectra of the two equities obtained from the MFDMA is much smaller than that from the MFDFA.
\end{abstract}

\begin{keyword}
 Econophysics; Stock and warrant; Intertrade duration; Correlation; Multifractal analysis
 \PACS 89.65.Gh, 89.75.Da, 02.50.-r, 89.90.+n, 05.65.+b
\end{keyword}

\end{frontmatter}

\section{Introduction}
\label{S1:Introduction}

The intertrade duration, defined as the waiting time between two consecutive transactions of an equity, is at the core of the autoregressive conditional duration (ACD) model \cite{Engle-Russell-1998-Em,Engle-Russell-1997-JEF,Engle-2000-Em}, which is deemed as the transaction-level version of the Nobel Prize winning autoregressive conditional heteroskedasticity (ARCH) model \cite{Engle-1982-Em}. With the availability of ultrahigh-frequency data, the statistical properties of the intertrade durations of some stocks have been investigated and some common and idiosyncratic patterns have been unveiled.

The most studied property is the distribution of intertrade durations of different stocks in different markets. There is conclusive evidence showing that the survival function of durations cannot be represented by a single exponential \cite{Scalas-Gorenflo-Luckock-Mainardi-Mantelli-Raberto-2004-QF,Scalas-Gorenflo-Luckock-Mainardi-Mantelli-Raberto-2005-FL,Scalas-Kaizoji-Kirchler-Huber-Tedeschi-2006-PA,Politi-Scalas-2007-PA}. Empirical analyses proposed diverse functional forms for the distributions, such as the Mittag-Leffler function \cite{Mainardi-Raberto-Gorenflo-Scalas-2000-PA,Sabatelli-Keating-Dudley-Richmond-2002-EPJB}, power laws \cite{Sabatelli-Keating-Dudley-Richmond-2002-EPJB,Yoon-Choi-Lee-Yum-Kim-2006-PA}, modified power laws \cite{Masoliver-Montero-Weiss-2003-PRE,Masoliver-Montero-Perello-Weiss-2006-JEBO}, stretched exponentials (or Weibulls) \cite{Bartiromo-2004-PRE,Raberto-Scalas-Mainardi-2002-PA,Ivanov-Yuen-Podobnik-Lee-2004-PRE,Vazquez-Oliveira-Dezso-Goh-Kondor-Barabasi-2006-PRE,Sazuka-2007-PA,Jiang-Chen-Zhou-2008-PA}, stretched exponentials followed power laws \cite{Kim-Yoon-2003-Fractals,Kim-Yoon-Kim-Lee-Scalas-2007-JKPS}. Recently, statistical tests were performed to illustrate that the waiting time distributions are neither exponentials nor power laws and the $q$-exponentials and the Weibulls are better candidates \cite{Poloti-Scalas-2008-PA}. For some stock markets, it has been found that the intertrade durations of different stocks exhibit a scaling behavior \cite{Ivanov-Yuen-Podobnik-Lee-2004-PRE,Jiang-Chen-Zhou-2008-PA} and the distribution is Weibull followed by a power law tail \cite{Jiang-Chen-Zhou-2008-PA}.

The long-term correlation or persistence in intertrade durations plays an important role in the ACD-type models \cite{Jasiak-1999-Finance}. In the econophysics community, the first research was conducted on 30 stocks listed on the New York Stock Exchange (NYSE) from January 1993 to December 1996 using the detrended fluctuation analysis (DFA) approach \cite{Ivanov-Yuen-Podobnik-Lee-2004-PRE}. In addition, crossover behaviors have been unveiled in the fluctuation scaling for many stocks on the NYSE, NASDAQ and Shenzhen Stock Exchange in different time periods \cite{Yuen-Ivanov-2005-XXX,Eisler-Kertesz-2006-EPJB,Jiang-Chen-Zhou-2009-PA}. Furthermore, the multifractal nature in the intertrade durations of German DAX stocks (from 28 November 1997 to 31 December 1999) and SZSE stocks (from 2 January 2003 to 31 December 2003) has been investigated and confirmed based on the multifractal detrended fluctuation analysis (MFDFA) approach \cite{Oswiecimka-Kwapien-Drozdz-2005-PA,Jiang-Chen-Zhou-2009-PA}.

The brief but complete review presented above shows that the statistical properties of intertrade durations have been studied only for stocks and relatively fewer studies have been conducted on the linear and nonlinear long-term correlations. In this work, we perform a comparative analysis of the statistical properties of the intertrade durations of a very liquid Chinese stock and its associated warrant. Special emphasis has been put on the correlations of the durations. Comparing with the data in Ref.~\cite{Jiang-Chen-Zhou-2008-PA,Jiang-Chen-Zhou-2009-PA}, the stock investigated in this work is listed on the Shanghai Stock Exchange (SHSE) and the data were recorded during a very bullish market phase. The results show that the two equities share very similar statistical properties.

\section{Data description}
\label{S1:Data}

In this work, we analyze the trading records of the Bao Steel stock and its warrant (a kind of call options) traded on the SHSE. The records started on 22 August 2005 and ended on 23 August 2006, containing 243 trading days. There are more than 25 million trades for the stock and about 3.8 million transactions for the warrant. Each data set records the relevant information of all the transactions. The time stamps are accurate to 1 second. For each equity, the time series of intertrade durations, denoted $\tau(i)$, is determined. In this procedure, only trades in the continuous double auction (from 9:30 a.m. to 11:30 a.m. and from 13:00 p.m. to 15:00 p.m.) are considered and the waiting times between the last trade in the morning and the first trade in the afternoon are excluded from the analysis. We find that 91.33\% intertrade durations of the stock are null and 82.44\% durations of the warrant are zero.

\section{Basic statistical properties}
\label{S1:BasicStatProp}

\subsection{Probability distribution}
\label{S2:PDF}

\begin{figure}[htb]
  \centering
  \includegraphics[width=8cm]{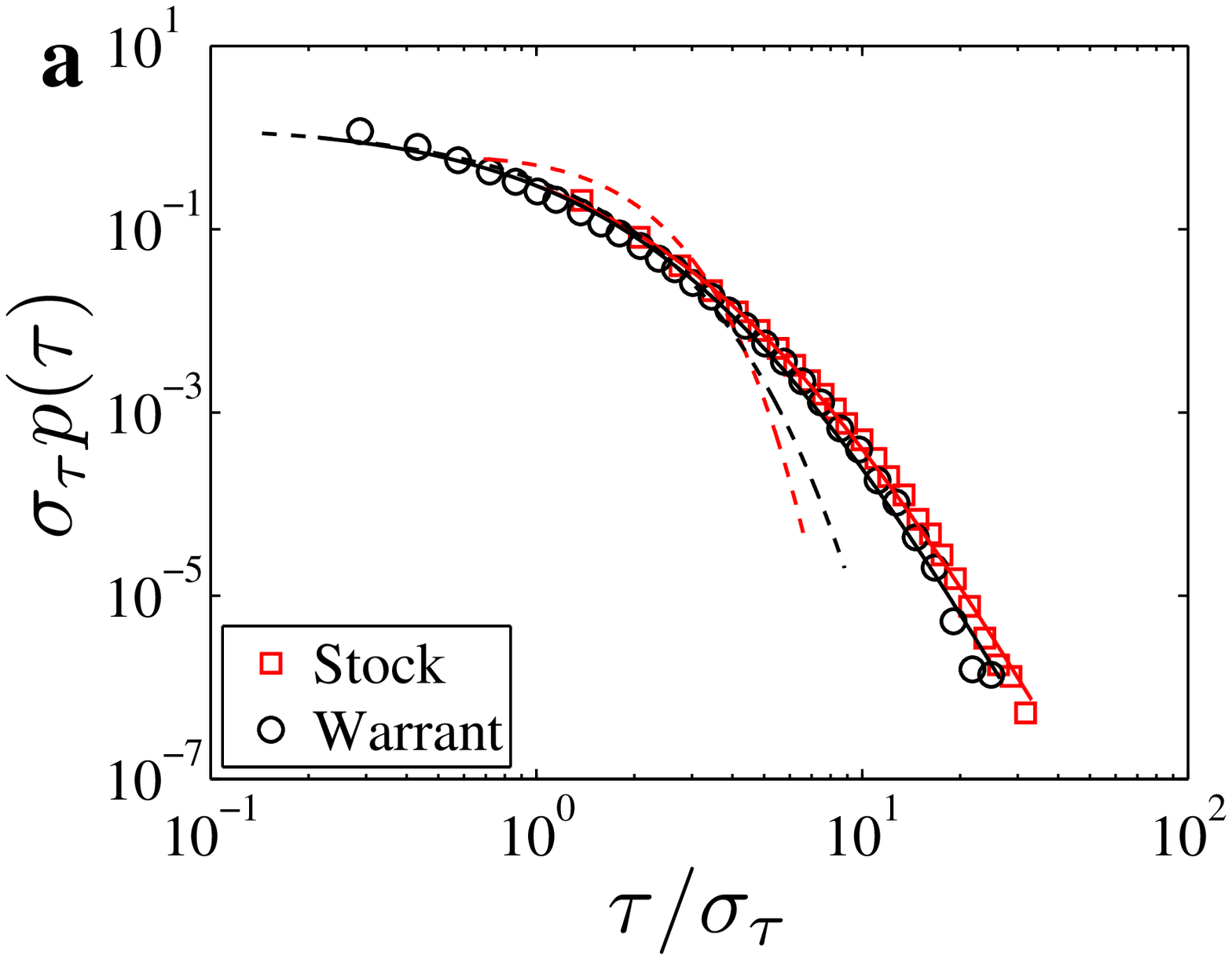}
  \includegraphics[width=8cm]{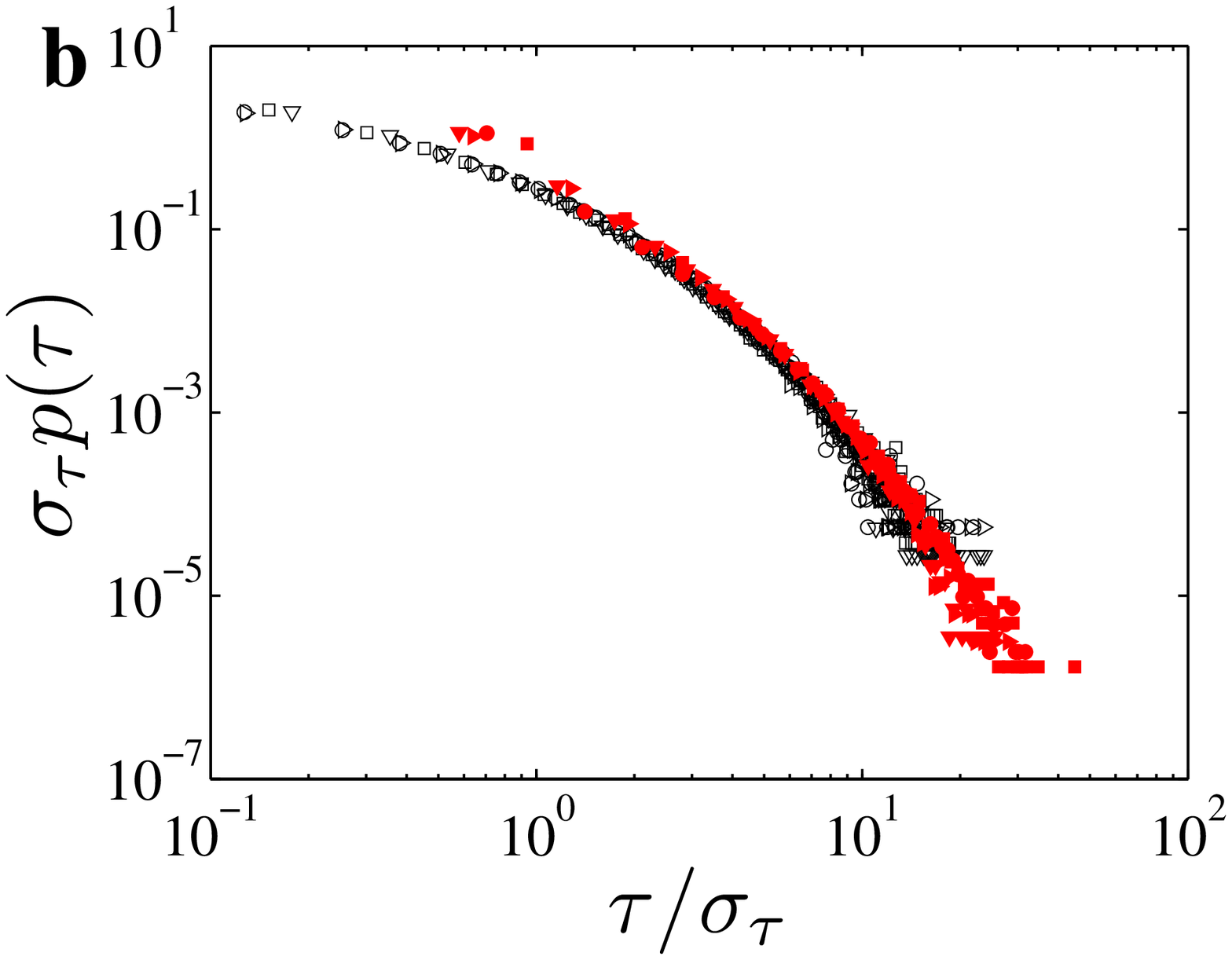}
  \caption{\label{Fig:ITD:PDF} (Color online) Empirical distributions of the scaled intertrade durations of the Bao Steel stock and its warrant in the whole time period (a) and in four successive time periods (b).}
\end{figure}

In order to compare the distributions of the two equities, we investigate the normalized durations $\tau/{\sigma_\tau}$, where $\sigma_\tau$ is the sample standard deviation of $\tau$ \cite{Jiang-Chen-Zhou-2008-PA}. Figure~\ref{Fig:ITD:PDF}(a) plots $\sigma_\tau p(\tau)$ as a function of $\tau/{\sigma_\tau}$ for the Bao Steel stock and its warrant. The most significant feature is that the scaled probability distributions of the stock and its warrant do not collapse onto a single curve, which is different from the scaling pattern for stocks \cite{Ivanov-Yuen-Podobnik-Lee-2004-PRE,Jiang-Chen-Zhou-2008-PA}.

Following Refs.~\cite{Poloti-Scalas-2008-PA,Jiang-Chen-Zhou-2008-PA}, we adopt the Weibull and the shifted power-law distributions to model the scaled intertrade durations, $g=\tau/\sigma_\tau$. The Weibull probability density $\rho_w(g)$ can be written as
\begin{equation}
 p_w(g)  = \alpha \beta g^{\beta-1} \exp(-\alpha g^{\beta}).
 \label{Eq:WeibellDensity}
\end{equation}
When $\beta=1$, $p_w(g)$ recovers the exponential distribution. When $0<\beta<1$, $p_w(g)$ is a stretched exponential or sub-exponential. When $\beta>1$, $p_w(g)$ is a super-exponential. We adopt the maximum likelihood estimator for model calibration and find that $\alpha=1.22$ and $\beta=1.41$ for the Bao Steel stock and $\alpha= 0.75$ and $\beta=0.97$ for the warrant. It means that the intertrade durations of the stock are sub-exponentially distributed, while the warrant has an exponential distribution in the intertrade durations. The resulting fits are illustrated in Fig.~\ref{Fig:ITD:PDF}(a) as dashed lines. It indicates that the Weibull distribution is not suitable for median and large durations.

The shifted power-law distribution is actually also known as the $q$-exponential distribution \cite{Tsallis-1988-JSP,Tsallis-Anteneodo-Borland-Osorio-2003-PA}, which is defined by
\begin{equation}
 p_q(g)  = a \left(g+g_0\right)^{-(\gamma+1)},
 \label{Eq:QDensity}
\end{equation}
where $g_0>0$. When $g\gg g_0$, we observe a power-law behavior in the tail with a tail exponent of $\gamma$. Nonlinear least-squares regressions give that $g_0= 4.09$ and $\gamma= 5.50$ for the stock and $g_0= 3.76$ and $\gamma= 5.70$ for the warrant. The resultant fits are illustrated in Fig.~\ref{Fig:ITD:PDF}(a) as continuous lines. We find that the tail behaviors of the durations are quite similar.

The difference in the intertrade duration distributions between the current stock and the SZSE stocks studied in Ref.~\cite{Jiang-Chen-Zhou-2008-PA} stems from the different time resolutions of the data. The super-exponential distribution describes the durations less than one second, while the sub-exponential distribution here concentrates on durations longer than one second. For median and large intertrade durations, the distributions are quite similar.

In order to investigate the stability of the distributions, we partition each data set into four segments with almost the same number of trading days. The empirical distributions in each time intervals for the two equities are plotted in Fig.~\ref{Fig:ITD:PDF}(b). It is obvious that the four distribution curves for each equity collapse excellently onto a single curve. It shows that the distribution is stable during different time intervals.

\subsection{Intraday pattern}
\label{S2:IntradayPattern}

The intertrade durations have been found to exhibit an inverse $U$-shaped intraday pattern in the NYSE market \cite{Engle-Russell-1998-Em,Hafner-2005-QF} and Russian stock market \cite{Anatolyev-Shakin-2007-AFE}, which indicates higher trading activity in the open and close than that in other time during each trading day. For the SZSE stocks traded in 2003, a different and more complex intraday pattern has been uncovered \cite{Jiang-Chen-Zhou-2009-PA}. It is interesting to check if the intraday pattern changed when new data are used.

We segment the continuous double auction of each trading day into 1440 successive 1-second intervals. For a given stock, we define an average intertrade duration for each interval as follows,
\begin{equation}
 \tau_{ij} = \frac{1}{N_{ij}} \sum_{k=1}^{N_{ij}} \tau_k,
  \label{Eq:tau:ij}
\end{equation}
where $\tau_{ij}$ is the average duration of the $j$-th interval in the $i$-th trading day, $N_{ij}$ represents the number of trades of the $j$-th interval in the $i$-th trading day. The average intertrade duration in the $j$-th time interval is calculated as follows,
\begin{equation}
 \langle \tau \rangle_j = \frac{1}{N_d} \sum_{i=1}^{N_d} \tau_{ij},
  \label{Eq:meanduration}
\end{equation}
where $N_d$ is the number of trading days. We note that the shape of the intraday pattern of each equity does not change no matter whether the durations equal to zero are considered or not.

\begin{figure}[htb]
  \centering
  \includegraphics[width=8cm]{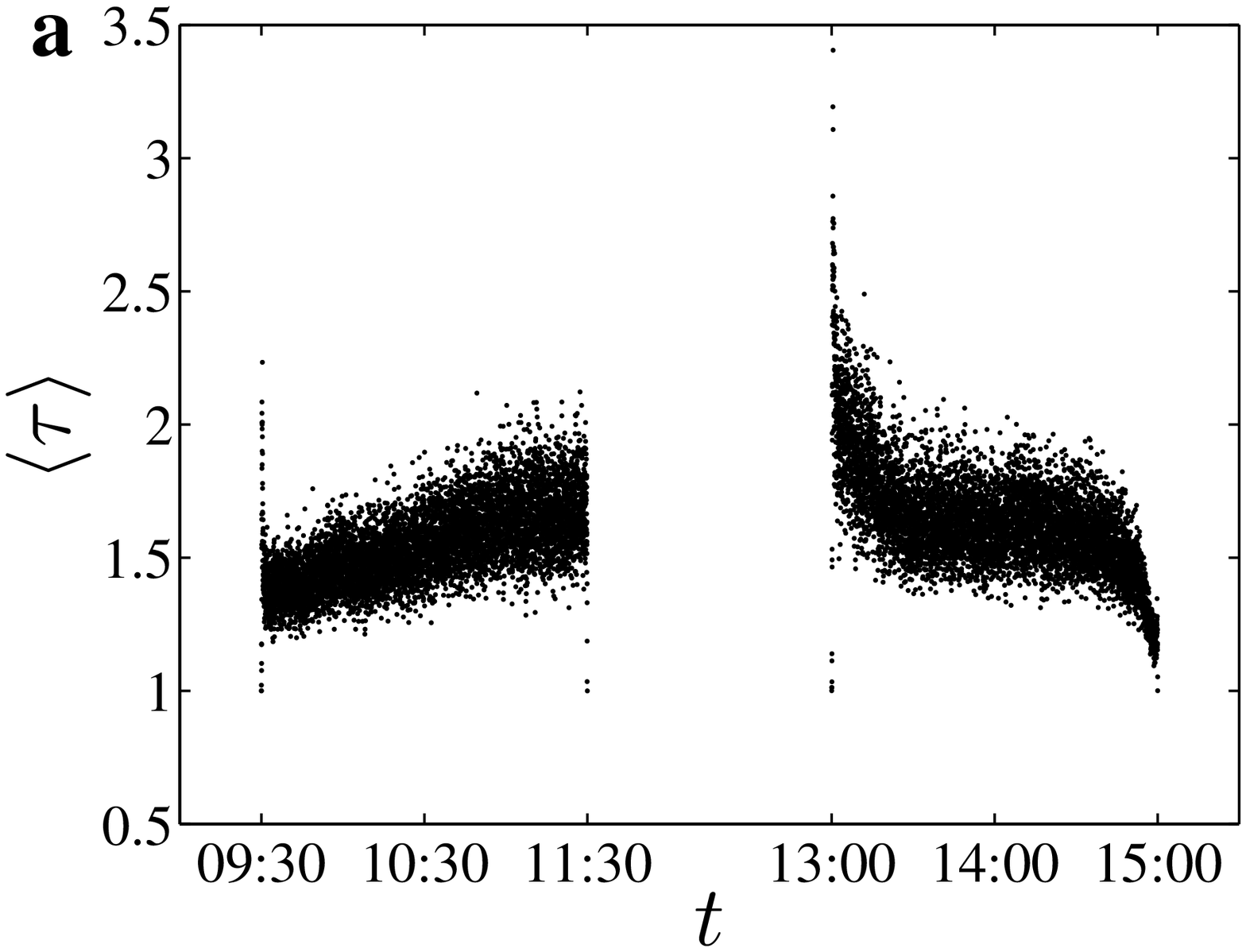}
  \includegraphics[width=8cm]{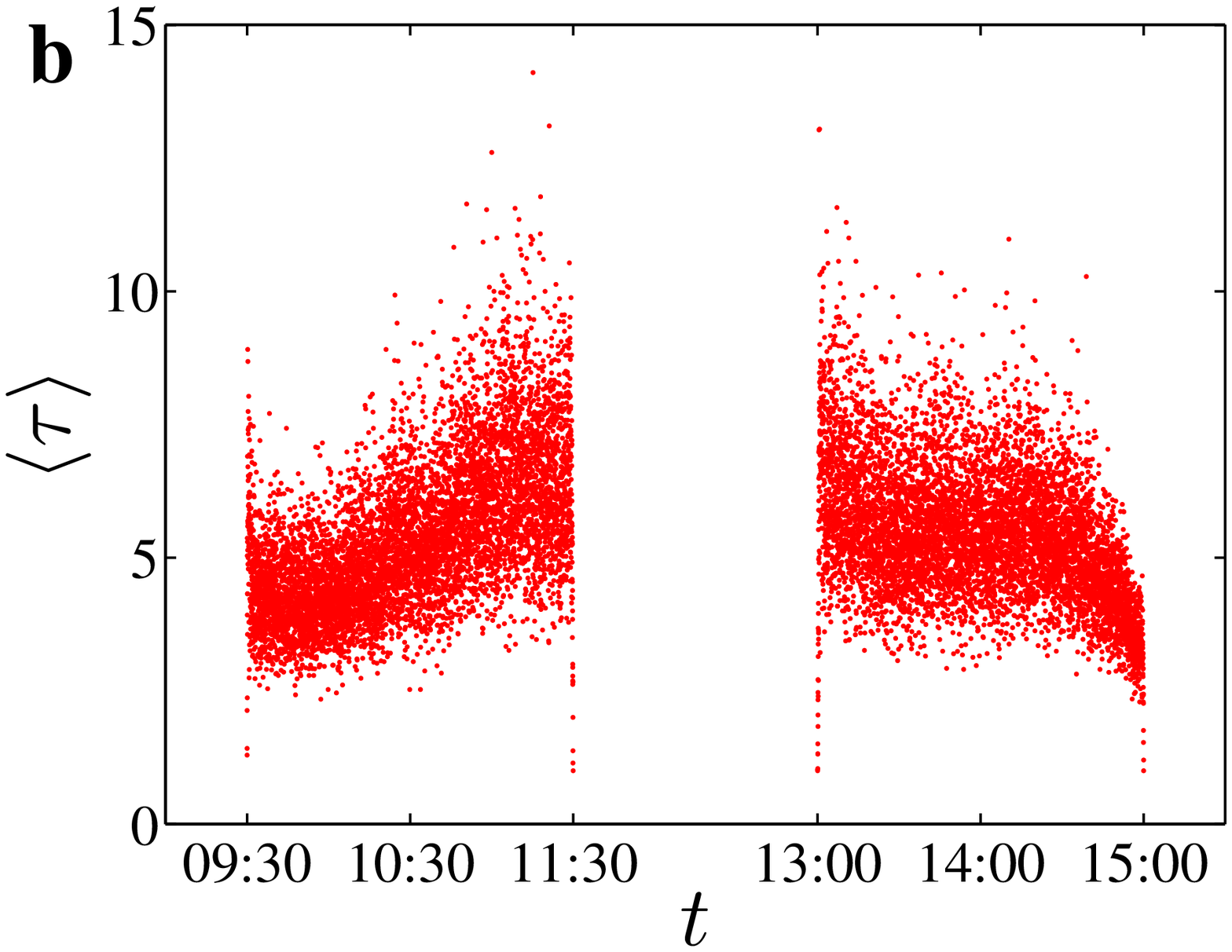}
  \caption{\label{Fig:ITD:IntradayPattern} (Color online) Intraday patterns of the intertrade durations in the continuous double auction for the Bao Steel stock (a) and its warrant (b) traded on the Shanghai Stock Exchange in 2005-2006.}
\end{figure}

The results are illustrated in Fig.~\ref{Fig:ITD:IntradayPattern}. It is interesting to observe that the two intraday patterns are qualitatively the same. The duration is small right after the opening of the two markets in the morning (9:30 a.m.) and in the afternoon (13:00 p.m.) and then increases fast to a large value followed by a relaxation in 10-20 seconds. When the time approaches 11:30 a.m. and 15:00 p.m., the duration sharply decreases to a small value in a few seconds. Excluding these short time intervals, the duration increases during the morning trading. In contrast, the duration decreases in the afternoon. These patterns are consistent with those for the SZSE stocks in 2003 \cite{Jiang-Chen-Zhou-2009-PA}. The higher resolution in Fig.~\ref{Fig:ITD:IntradayPattern} allows us to uncover the fine structure in the intraday patterns.

\section{Long-term correlations}
\label{S1:Correlation}

There are numerous methods proposed for the detection of long-term correlations in time series \cite{Taqqu-Teverovsky-Willinger-1995-Fractals,Montanari-Taqqu-Teverovsky-1999-MCM}. For instance, researchers frequently applies the rescaled range analysis \cite{Hurst-1951-TASCE,Mandelbrot-Ness-1968-SIAMR,Mandelbrot-Wallis-1969a-WRR,Mandelbrot-Wallis-1969b-WRR,Mandelbrot-Wallis-1969c-WRR,Mandelbrot-Wallis-1969d-WRR}, the fluctuation analysis \cite{Peng-Buldyrev-Goldberger-Havlin-Sciortino-Simons-Stanley-1992-Nature}, the detrended fluctuation analysis (DFA)
\cite{Peng-Buldyrev-Havlin-Simons-Stanley-Goldberger-1994-PRE,Hu-Ivanov-Chen-Carpena-Stanley-2001-PRE,Kantelhardt-KoscielnyBunde-Rego-Havlin-Bunde-2001-PA}, the wavelet transform module maxima (WTMM) approach \cite{Holschneider-1988-JSP,Bacry-Muzy-Arneodo-1993-JSP}, and the detrending moving average (DMA) algorithm \cite{Vandewalle-Ausloos-1998-PRE,Alessio-Carbone-Castelli-Frappietro-2002-EPJB,Carbone-Castelli-Stanley-2004-PRE}, to name a few. In order to investigate the possible long-term correlations in the intertrade durations of the Bao Steel stock and its warrant, we adopt the DFA and the DMA algorithms for comparison.

\subsection{DFA and DMA algorithms}

The idea of DFA was invented originally to investigate the long-range dependence in coding and noncoding DNA nucleotide sequences \cite{Peng-Buldyrev-Havlin-Simons-Stanley-Goldberger-1994-PRE}. Due to its efficiency and simplicity in implementation, the DFA is now becoming the most important method in the field \cite{Zhou-2007}. The DFA algorithm can be implemented as follows.

For a given intertrade duration series $\{\tau(i)|i =1,2,\cdots,N\}$, define the cumulative summation series $y(i)$ as follows,
\begin{equation}
  y(i) = \sum_{j=1}^{i} \left[\tau(i)-\langle{\tau}\rangle\right],~~i = 1, 2, \cdots, N,
  \label{Eq:cumsum}
\end{equation}
where $\langle{\tau}\rangle$ is the sample mean of the $\tau(i)$ series. The series $y$ is covered by $N_s$ disjoint boxes with the same size $s$. When the whole series $y(i)$ cannot be completely covered by $N_s$ boxes, we can utilize $2N_s$ boxes to cover the series from both ends of the series. In each box, a polynomial trend function $g(i)$ of the sub-series is determined. The residuals are calculated by
\begin{equation}
  \epsilon(i) = y(i)-g(i),
  \label{Eq:DFA:epsilon}
\end{equation}
and the local detrended fluctuation function $f_v(s)$ in the $v$-th box is defined as the r.m.s. of the fitting residuals:
\begin{equation}
  \left[f_v(s)\right]^2 = \frac{1}{s}\sum_{i=(v-1)s+1}^{vs} \left[\epsilon(i)\right]^2~.
  \label{Eq:fv:s}
\end{equation}
The overall detrended fluctuation is estimated as follows
\begin{equation}
  \left[F(s)\right]^2 = \frac{1}{N_s}\sum_{i=1}^{N_s} \left[f_v(s)\right]^2.
  \label{Eq:F2:s}
\end{equation}
As the box size $s$ varies in the range of $[20,N/4]$, one can determine the power law relationship between the overall fluctuation function $F(s)$ and the box size $s$,
\begin{equation}
  F(s) \sim s^H,
  \label{Eq:Hurst}
\end{equation}
where $H$ signifies the DFA scaling exponent, which is related to the power spectrum exponent $\eta$ by $\eta = 2H-1$ and to the autocorrelation exponent $\gamma$ by $\gamma = 2-2H$ \cite{Talkner-Weber-2000-PRE,Heneghan-McDarby-2000-PRE}.

The DMA approach is based on the moving average (MA) or mobile average technique \cite{Carbone-2009-IEEE}, which was proposed to estimate the Hurst exponent of self-affinity signals \cite{Vandewalle-Ausloos-1998-PRE} and further developed to the detrending moving average algorithms by considering the second-order difference between the original signal and its moving average function \cite{Alessio-Carbone-Castelli-Frappietro-2002-EPJB}. Because the DMA method can also be easily implemented to estimate the correlation properties of non-stationary series without any assumption, it is widely applied to the analysis of real-world time series \cite{Carbone-Castelli-2003-SPIE,Carbone-Castelli-Stanley-2004-PA,Carbone-Castelli-Stanley-2004-PRE,Varotsos-Sarlis-Tanaka-Skordas-2005-PRE,Serletis-Rosenberg-2007-PA,Arianos-Carbone-2007-PA,Matsushita-Gleria-Figueiredo-Silva-2007-PLA,Serletis-Rosenberg-2009-CSF} and synthetic signals as well \cite{Serletis-2008-CSF}. Extensive numerical experiments unveil that the performance of the DMA method is comparable to the DFA method with slightly different priorities under different situations \cite{Xu-Ivanov-Hu-Chen-Carbone-Stanley-2005-PRE,Bashan-Bartsch-Kantelhardt-Havlin-2008-PA}.

The main difference between the DFA and DMA algorithms is about the detrending procedure. In the DMA approach, one calculates the moving average function $\widetilde{y}(i)$ in a moving window \cite{Arianos-Carbone-2007-PA},
\begin{equation}
\widetilde{y}(i)=\frac{1}{s}\sum_{k=-\lfloor(s-1)\theta\rfloor}^{\lceil(s-1)(1-\theta)\rceil}y(i-k),
\label{Eq:1ddma:y1}
\end{equation}
where $s$ is the window size, $\lfloor{x}\rfloor$ is the largest integer not greater than $x$, $\lceil{x}\rceil$ is the smallest integer not smaller than $x$, and $\theta$ is the position parameter with the value varying in the range $[0,1]$. Hence, the moving average function considers $\lceil(s-1)(1-\theta)\rceil$ data points in the past and $\lfloor(s-1)\theta\rfloor$ points in the future. We consider three special cases in this paper. The first case $\theta=0$ refers to the backward moving average \cite{Xu-Ivanov-Hu-Chen-Carbone-Stanley-2005-PRE}, in which the moving average
function $\widetilde{y}(i)$ is calculated over all the past $s-1$ data points of the signal. The second case $\theta=0.5$ corresponds to the centered moving average \cite{Xu-Ivanov-Hu-Chen-Carbone-Stanley-2005-PRE}, where $\widetilde{y}(i)$ contains half past and half future information in each window. The third case $\theta=1$ is called the forward moving average, where $\widetilde{y}(i)$ considers the trend of $s-1$ data points in the future. The residual sequence $\epsilon(i)$ is obtained by removing the moving average function $\widetilde{y}(i)$ from $y(i)$:
\begin{equation}
  \epsilon(i)=y(i)-\widetilde{y}(i),
  \label{Eq:1ddma:epsilon}
\end{equation}
where $s-\lfloor(s-1)\theta\rfloor\leqslant{i}\leqslant{N-\lfloor(s-1)\theta\rfloor}$. We can then calculate the overall fluctuation function $F(s)$ as in Eq.~(\ref{Eq:F2:s}) and calculate the DMA scaling exponent through investigating the dependence between $F(s)$ and $s$.

\subsection{Results}

\begin{figure}[htb]
  \centering
  \includegraphics[width=8cm]{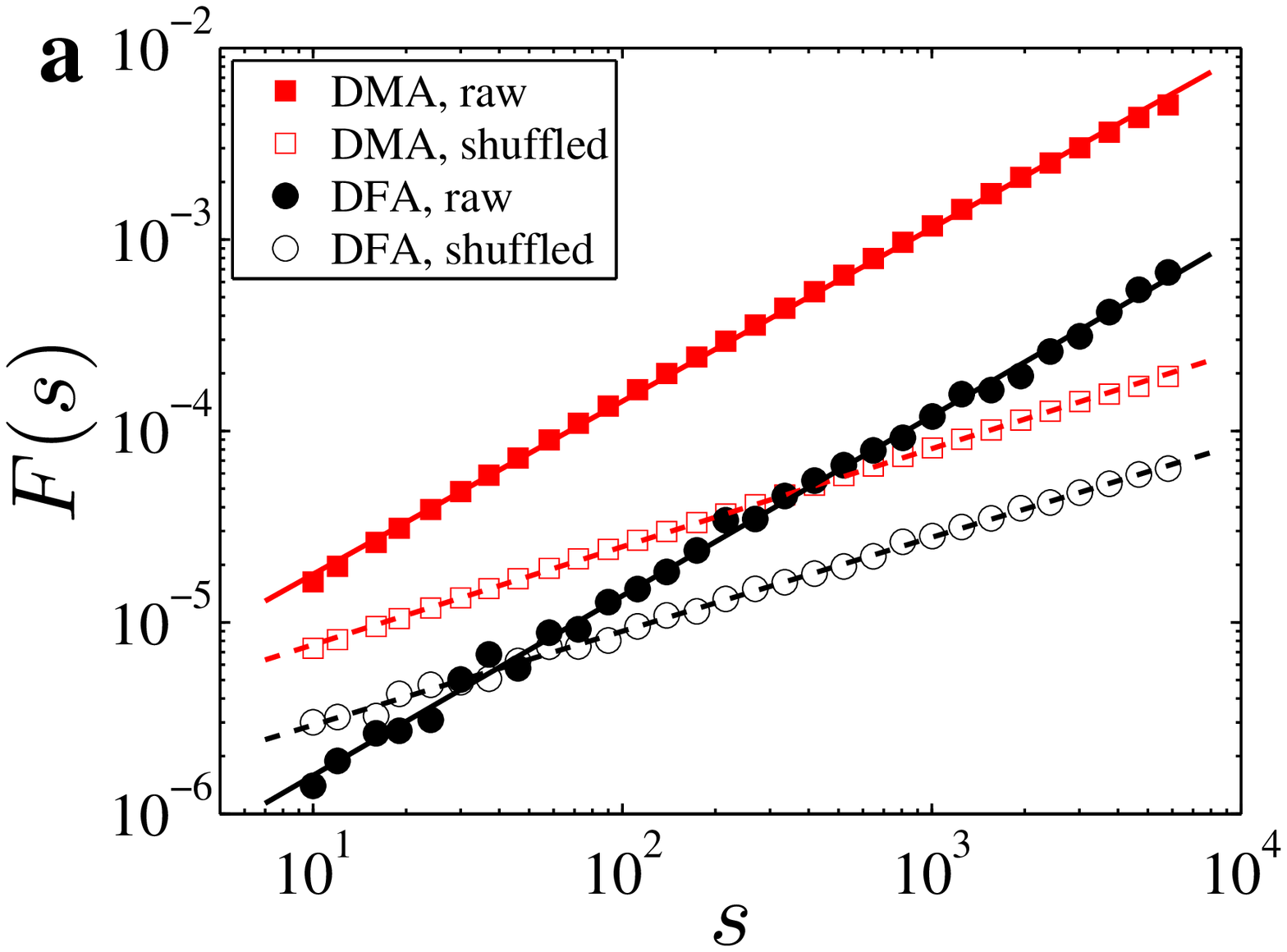}
  \includegraphics[width=8cm]{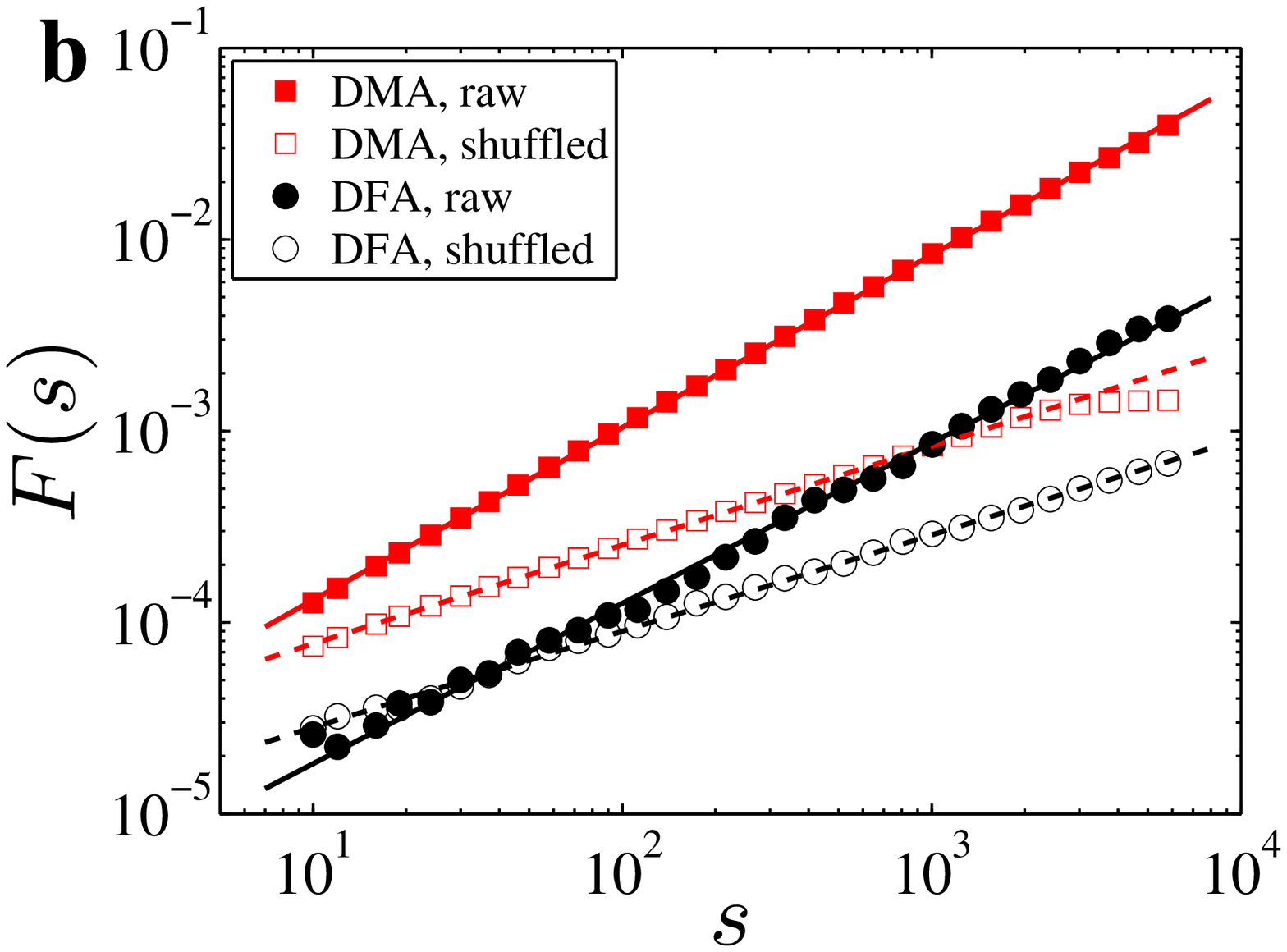}
  \caption{\label{Fig:BG:ITD:Hurst} (Color online) Detection of long-term correlations in the intertrade duration time series of the Bao Steel stock (a) and its warrant (b) using the detrended fluctuation analysis and the detrending moving average analysis.}
\end{figure}

Figure \ref{Fig:BG:ITD:Hurst}(a) shows respectively the dependence of the DFA and DMA fluctuations $F(s)$ of the intertrade durations of the Bao Steel stock as a function of the time scale $s$ in double logarithmic coordinates. Nice power-law scaling relations are observed, where the scaling ranges span over three orders of magnitude. We obtain that the DMA scaling exponent is $H_{\rm{s,DMA}}=0.904\pm0.008$ and the DFA scaling exponent is $H_{\rm{s,DFA}}=0.939\pm0.019$, where the error bars are obtained at the significance level of 5\%. We also perform DMA and DFA analysis on the shuffled time series of the intertrade durations and present the results in Fig.~\ref{Fig:BG:ITD:Hurst}(a). We find that, for the shuffled data of the stock, the DMA exponent is $H_{\rm{s',DMA}}=0.512\pm0.004$ and the DFA exponent is $H_{\rm{s',DFA}}=0.491\pm0.008$. According to Fig.~\ref{Fig:BG:ITD:Hurst}(a), it is clear that the intertrade durations of the Bao Steel stock are long-term correlated, while the shuffled data are uncorrelated.

Figure \ref{Fig:BG:ITD:Hurst}(b) shows respectively the dependence of the DFA and DMA fluctuations $F(s)$ of the intertrade durations of the Bao Steel warrant as a function of the time scale $s$ in double logarithmic coordinates. We also observe nice power-law scaling relations with wide scaling ranges spanning over three orders of magnitude. We obtain that the DMA scaling exponent is $H_{\rm{w,DMA}}=0.902\pm0.003$ and the DFA scaling exponent is $H_{\rm{w,DFA}}=0.838\pm0.017$. For the shuffled data of the warrant, we find that the DMA exponent is $H_{\rm{w',DMA}}=0.516\pm0.003$ and the DFA exponent is $H_{\rm{w',DFA}}=0.503\pm0.008$. According to Fig.~\ref{Fig:BG:ITD:Hurst}(b), it is clear that the intertrade durations of the Bao Steel warrant are long-term correlated, while the shuffled data are uncorrelated.

An interesting feature is observed in Fig.~\ref{Fig:BG:ITD:Hurst} that the DMA fluctuation function $F(s)$ has a better power-law relation with $s$. Specifically, we calculate the r.m.s of the fitting errors of the eight lines and obtain that $E_{\rm{s,DMA}}=0.0397$, $E_{\rm{s,DFA}}=0.0910$, $E_{\rm{s',DMA}}=0.0177$, and $E_{\rm{s',DFA}}=0.0410$ for the Bao Steel stock and $E_{\rm{w,DMA}}=0.0143$, $E_{\rm{w,DFA}}=0.0785$, $E_{\rm{w',DMA}}=0.0117$, and $E_{\rm{w',DFA}}=0.0378$ for the Bao Steel warrant. It is clear that
\begin{equation}
 E_{\rm{e,DMA}} < E_{\rm{e,DFA}},
 \label{Eq:E:DMA:DFA}
\end{equation}
where the subscript ``$\rm{e}$'' stands for $\rm{s}$, $\rm{s'}$, $\rm{w}$, and $\rm{w'}$, respectively. This empirical observation is consistent with the numerical results in Ref.~\cite{Gu-Zhou-2010-PRE}. However, we stress that the DMA and the DFA give qualitatively the same results.

In the previous studies, two scaling ranges in the fluctuation function have been observed for stocks in different markets \cite{Ivanov-Yuen-Podobnik-Lee-2004-PRE,Yuen-Ivanov-2005-XXX,Eisler-Kertesz-2006-EPJB,Jiang-Chen-Zhou-2009-PA}. This crossover phenomenon disappears in Fig.~\ref{Fig:BG:ITD:Hurst}, whose cause is unknown yet.

\section{Multifractal analysis}
\label{S1:MF}

Since the seminal works in the 1990's \cite{Ghashghaie-Breymann-Peinke-Talkner-Dodge-1996-Nature,Mantegna-Stanley-1996-Nature,Vandewalle-Ausloos-1998-EPJB,Mandelbrot-1999-SA}, there have been a wealth of studies showing that financial markets exhibit multifractal nature \cite{Ivanova-Ausloos-1999-EPJB,Schmitt-Schertzer-Lovejoy-1999-ASMDA,Schmitt-Schertzer-Lovejoy-2000-IJTAF,
Calvet-Fisher-2002-RES,Ausloos-Ivanova-2002-CPC,Kantelhardt-Zschiegner-KoscielnyBunde-Havlin-Bunde-Stanley-2002-PA,Struzik-Siebes-2002-PA,Gorski-Drozdz-Speth-2002-PA,AlvarezRamirez-Cisneros-IbarraValdez-Soriano-2002-PA,
Balcilar-2003-EMFT,Matia-Ashkenazy-Stanley-2003-EPL,Turiel-Perez-Vicente-2003-PA,
Lee-Lee-2005a-JKPS,Lee-Lee-2005b-JKPS,Eisler-Kertesz-Yook-Barabasi-2005-EPL,Kwapien-Oswiecimka-Drozdz-2005-PA,Oswiecimka-Kwapien-Drozdz-2005-PA,Turiel-Perez-Vicente-2005-PA,Oswiecimka-Kwapien-Drozdz-Rak-2005-APPB,
Lee-Lee-Rikvold-2006-PA,
Eisler-Kertesz-2007-EPL,Jiang-Guo-Zhou-2007-EPJB,Jiang-Zhou-2007-PA,Jiang-Ma-Cai-2007-PA,Lee-Lee-2007-PA,Lim-Kim-Lee-Kim-Lee-2007-PA,
Jiang-Zhou-2008a-PA,Su-Wang-Huang-2009-JKPS,Jiang-Chen-Zhou-2009-PA,Su-Wang-2009-JKPS,Chen-He-2010-PA,He-Chen-2010a-PA,He-Chen-2010b-PA}.
The methods adopted in these studies include the wavelet transform module maxima approach
\cite{Muzy-Bacry-Arneodo-1991-PRL,Muzy-Bacry-Arneodo-1993-PRE,Muzy-Bacry-Arneodo-1994-IJBC}, the multifractal detrended fluctuation analysis (MFDFA)
\cite{Kantelhardt-Zschiegner-KoscielnyBunde-Havlin-Bunde-Stanley-2002-PA}, the partition function and structure function approach \cite{Frisch-1996}, and so on. Here we adopt the MFDFA algorithm and the multifractal detrending moving average (MFDMA) algorithm proposed recently \cite{Gu-Zhou-2010-PRE}.

\subsection{MFDFA and MFDMA algorithms}

The MFDFA algorithm is a generalization of the DFA algorithm \cite{Kantelhardt-Zschiegner-KoscielnyBunde-Havlin-Bunde-Stanley-2002-PA}. It is necessary to note that the MFDFA approach was independently invented by two other groups earlier but with much less visibility in the community \cite{CastroESilva-Moreira-1997-PA,Talkner-Weber-2001-JGR}. On the other hand, the MFDMA method is an extension of the DMA technique \cite{Gu-Zhou-2010-PRE}, which is very similar to the MFDFA except that they use different detrending procedures as described in Section \ref{S1:Correlation}.

For both MFDFA and MFDMA, the $q$th-order overall detrended fluctuation in Eq.~(\ref{Eq:F2:s}) is generalized to the following form
\begin{equation}
  F_q(s) = \left\{\frac{1}{N_s}\sum_{v=1}^{N_s} {F_v^q(s)}\right\}^{\frac{1}{q}},
  \label{Eq:Fqs}
\end{equation}
where $q$ can take any real value except for $q=0$. When $q=0$,
we have
\begin{equation}
  \ln[F_0(s)] = \frac{1}{N_s}\sum_{v=1}^{N_s}{\ln[F_v(s)]},
  \label{Eq:Fq0}
\end{equation}
according to L'H\^{o}spital's rule. Varying the values of segment size $s$, we can determine the power-law relation between the function $F_q(s)$ and the size scale $s$,
\begin{equation}
  F_q(s) \sim {s}^{h(q)}.
  \label{Eq:hq}
\end{equation}

According to the standard multifractal formalism, the multifractal scaling exponent $\tau(q)$ can be used to characterize the multifractal nature, which reads
\begin{equation}
\tau(q)=qh(q)-D_f,
\label{Eq:tau:hq}
\end{equation}
where $D_f$ is the fractal dimension of the geometric support of the multifractal measure \cite{Kantelhardt-Zschiegner-KoscielnyBunde-Havlin-Bunde-Stanley-2002-PA}. For time series analysis, we have $D_f=1$. If the scaling exponent function $\tau(q)$ is a nonlinear function of $q$, the time series is regarded to have multifractal nature. It is easy to obtain the singularity strength function $\alpha(q)$ and the multifractal spectrum $f(\alpha)$ through the Legendre transform \cite{Halsey-Jensen-Kadanoff-Procaccia-Shraiman-1986-PRA}
\begin{equation}
    \left\{
    \begin{array}{ll}
        \alpha(q)={\rm{d}}\tau(q)/{\rm{d}}q\\
        f(q)=q{\alpha}-{\tau}(q)
    \end{array}
    \right..
\label{Eq:f:alpha:tau}
\end{equation}

\subsection{Results}

\begin{figure}[htb]
  \centering
  \includegraphics[width=8cm]{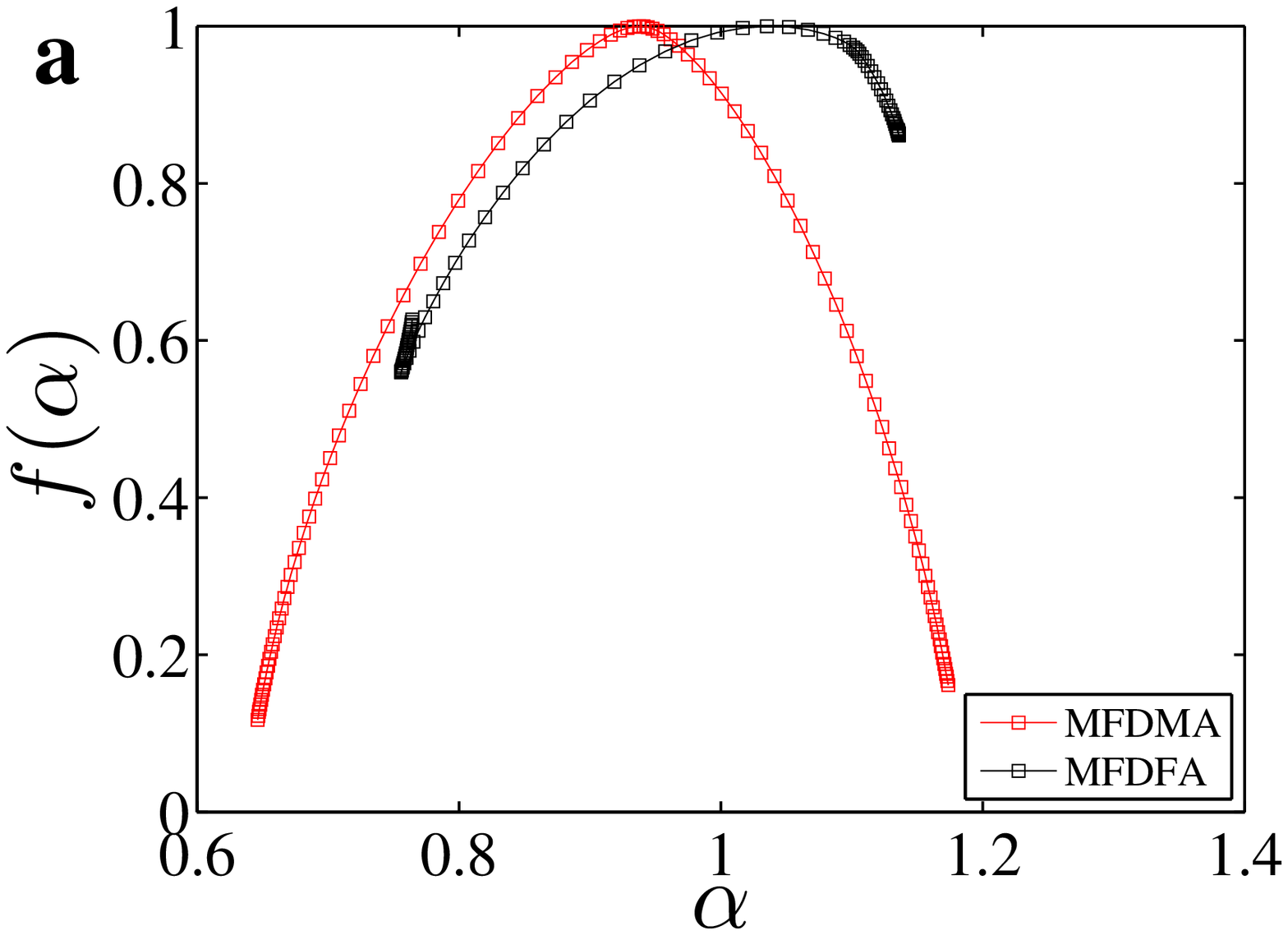}
  \includegraphics[width=8cm]{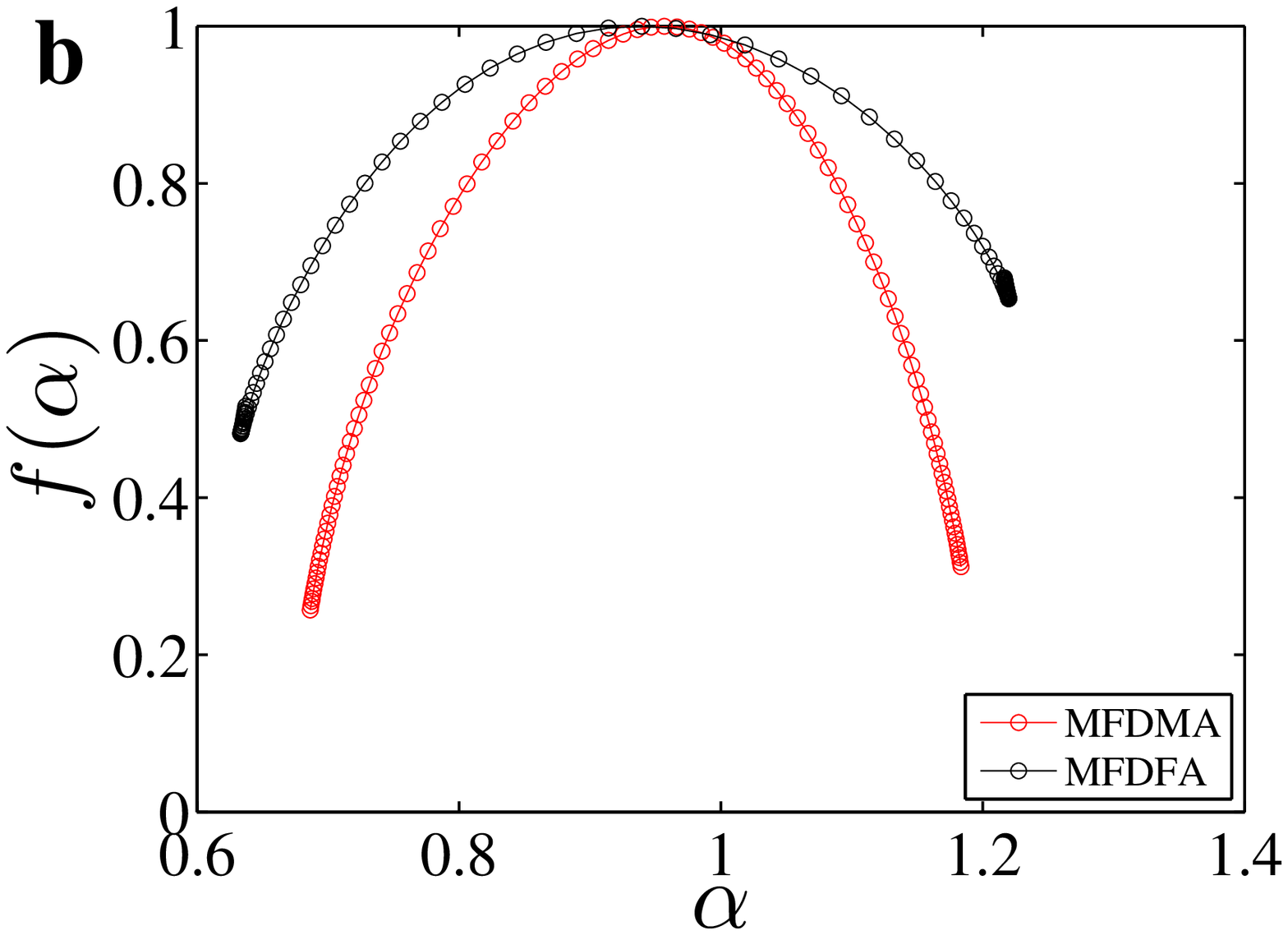}
  \caption{\label{Fig:BG:ITD:MF} (Color online) Multifractal spectra of the intertrade duration time series of the Bao Steel stock (a) and its warrant (b) obtained according to the detrended fluctuation analysis and the detrending moving average analysis.}
\end{figure}

For the SZSE stocks in 2003, a crossover phenomenon has been reported in the detrended fluctuation functions \cite{Jiang-Chen-Zhou-2009-PA}, which is not observed in the current case for the MFDFA and MFDMA fluctuation functions. The four multifractal spectra of the intertrade duration time series of the Bao Steel stock and its warrant are illustrated in Fig.~\ref{Fig:BG:ITD:MF}. It is found that both time series exhibit multifractal nature.

For each equity, the multifractal spectra extracted from the two methods are significantly different. The two multifractal spectra for the two equities obtained from the MFDMA approach are quantitatively similar, while other two spectra from the MFDFA method differ significantly.

\section{Conclusion}

In summary, we have performed a comparative study to investigate the statistical properties of the intertrade durations of a very liquid Chinese stock and its warrant using ultrahigh-frequency data. In a nutshell, we found that the statistical properties are qualitatively similar. The distributions of the two equities can be better described by the shifted power-law form than the Weibull and their scaled distributions do not collapse into a single curve. Both detrended fluctuation analysis and detrending moving average analysis showed that the 1-min intertrade duration time series of the two equities are strongly correlated. In addition, the multifractal nature has been confirmed in the intertrade durations based on the MFDFA and MFDMA approaches. Contrary to the previous works, we found no crossover behaviors in the fluctuation functions.

\bigskip
{\textbf{Acknowledgments:}}

We thank Zhi-Qiang Jiang for preprocessing the data. We acknowledge financial supports from the Natural Science Foundation of China (Grant no. 70803010), the Program for New Century Excellent Talents in University (Grant No. NCET-07-0288) and the Fundamental Research Funds for the Central Universities.

\bibliography{E:/Papers/Auxiliary/Bibliography}

\end{document}